\documentstyle[prd,aps,amsfonts,eqsecnum]{revtex}

\begin{document}

\draft

\title{Renormalized black hole entropy in anti-de Sitter space
via the ``brick wall'' method}

\author{Elizabeth Winstanley 
\thanks{e-mail: E.Winstanley@sheffield.ac.uk}}

\address{Department of Applied Mathematics,
The University of Sheffield,\\
Hicks Building,
Hounsfield Road,
Sheffield.
S3 7RH
U. K. }

\date{\today}

\maketitle

\begin{abstract}
We consider the entropy of a quantum scalar
field on a background black hole geometry
in asymptotically anti-de Sitter space-time,
using the ``brick wall'' approach.
In anti-de Sitter space, the theory has
no infra-red divergences, and all ultra-violet
divergences can be absorbed into a renormalization
of the coupling constants in the one-loop
effective gravitational Lagrangian.
We then calculate the finite renormalized entropy
for the Schwarzschild-anti-de Sitter and extremal
Reissner-Nordstr\"om-anti-de Sitter black holes,
and show that, at least for large black holes,
the entropy is entirely accounted for by the
one-loop Lagrangian, apart possibly from terms
proportional to the logarithm of the 
event horizon radius.  
For small black holes, there are indications
that non-perturbative quantum gravity
effects become important. 
\end{abstract}


\section{Introduction}
\label{sec:intro}
The origin and understanding of black entropy has been a 
fruitful area of research for nearly three decades,
since the original proposal by Bekenstein~\cite{bekenstein73} 
that the entropy of a black hole should be proportional to
the area of the event horizon.
Since then, there has been a large body of work 
attempting to understand the microscopic origin
of this black hole entropy.
In view of this sizable literature on the subject,
we shall not be able to give a complete account
of all developments, nor more than a small selection
of relevant references.
In this article we shall focus on a semi-classical
approach, in which the black hole geometry is 
considered to be a fixed classical background on 
which quantum fields propagate.  
This was the approach taken by 't Hooft~\cite{thooft85},
who considered the entropy of a thermal gas of
particles outside the event horizon of a 
Schwarzschild black hole, using
the WKB approximation.
The calculation involves divergences coming from
the number of modes close to the event horizon,
which were regulated by using a ``brick wall'', namely
a cut-off just outside the event horizon.

In the original paper~\cite{thooft85}, 't Hooft calculated
the leading order divergences in the entropy, and
found that they were proportional to the area of
the event horizon multiplied by ${\tilde {\epsilon }}^{-2}$,
where ${\tilde {\epsilon }}$ is the proper distance
of the ``brick wall'' from the event horizon
(see also~\cite{mann92} for early work on this topic,
extending 't Hooft's original results to more general
black holes and dimensions other than four).
It was subsequently suggested~\cite{susskind94,barbon95}
that this divergence could be absorbed in a 
renormalization of Newton's constant.
The next stage was to consider the next-to-leading order
divergences, which turn out to be proportional to 
$\log {\tilde {\epsilon }}$~\cite{alwis95,solodukhin95}.
After some discussion of the interpretation 
of these terms~\cite{solodukhin95,mavromatos96}, it was agreed
that they can be absorbed into a renormalization
of the coupling constants in the one-loop
effective gravitational Lagrangian, which 
contains terms quadratic in the 
curvatures~\cite{demers95,fursaev96,shimomura00}.
This then accounts for all the divergent terms
in the entropy, so the remaining quantity is
finite.
However, this finite quantity has not 
received much attention and it is the 
focus of this paper (see~\cite{solodukhin96} for
work in this area in two dimensions).

Subsequently this ``brick wall'' scenario has been
studied by many authors, using various approaches,
and agreement has been obtained on the
renormalization of the quantum field entropy
using the one-loop effective Lagrangian.
For example, the ``brick wall'' approach
has been regularized using 
Pauli-Villars regularization~\cite{demers95,kim96},
the renormalization has been confirmed
via a conical singularity approach~\cite{fursaev96}
and the result has been extended to 
fields with non-zero spin~\cite{shimomura00}.

We should, at this stage, mention that there
has been some discussion in the literature
as to what this ``brick wall'' model
represents physically.
One objection to the model 
might be that the introduction of the ``brick wall''
just outside the event horizon is
somewhat unphysical.
Fortunately, the results outlined
above mean that the ``brick wall''
is simply a useful mathematical tool
to regularize the theory, and 
it can be completely removed by
renormalization.
The status of the ``brick wall'' model
(with the ``brick wall'' in place)
has been put on a firmer footing 
recently~\cite{mukhoyama98}, 
the problem having been considered
in a similar vein previously~\cite{frolov96}.
Mukhoyama and Israel~\cite{mukhoyama98} showed that the quantum
state studied in the ``brick wall''
model represents a thermally excited
state above the Boulware vacuum, which 
is the ground state in this scenario.
In other words, expectation 
values of operators calculated for
this state are the differences
between expectation values in the
Hartle-Hawking and Boulware states.
This accounts for the divergent contribution
to the entropy from close to the event horizon.
The difference between the two situations
is that, in our case, the divergences can
be renormalized away, whereas the divergences
in the Boulware state expectation values
(such as the expectation value of
the stress tensor)
arise in quantities which have already 
been renormalized.
However, this work leads us to conclude
that the ``brick wall'' model we shall
study in this article is a useful and 
physically reasonable aid to understanding
black hole entropy.

In this article we will use 't Hooft's 
original ``brick wall'' approach, and
we begin in section~\ref{sec:renorm} 
by studying the divergent terms
coming from close to the event horizon.
We consider a general spherically symmetric
black hole in four dimensions, 
and a minimally coupled scalar field.
The effects of rotation~\cite{ho97},
higher dimensions~\cite{mann92,kim97} 
and non-minimal coupling~\cite{solodukhin97}
have been considered elsewhere.
We confirm that
all the divergences can indeed
be renormalized away.
We use this approach because it is
the most technically straightforward,
especially since the divergences
are under control.
We consider black holes in geometries
with a negative cosmological constant, for
two reasons:
\begin{enumerate}
\item
Firstly, the introduction of a negative
cosmological constant regularizes the 
contribution to the entropy from infinity.
Although in asymptotically flat space
this contribution is well understood as the 
entropy of a quantum field in flat space, 
it will simplify our calculations and 
interpretation if this term is absent.
\item
Secondly, black holes in asymptotically anti-de
Sitter (adS) space can be in stable equilibrium
with a thermal heat bath of radiation at the 
Hawking temperature, provided that
the black hole is sufficiently large~\cite{hawking83}.
Since we are considering a quantum field in a thermal
state, it seems most appropriate to consider
black holes for which this configuration is 
stable.
\end{enumerate}

Our focus in this paper is non-extremal black holes.
The application of the ``brick wall'' model
to extremal black holes has been the subject of
some controversy~\cite{alwis95,kim96,ghosh94,cai96,ghosh95,ghosh97,mann98}.
In common with other approaches to black hole
entropy, there are two routes to
studying extremal black holes.
Firstly, one can take the ``extremalization after
quantization''~\cite{ghosh97} approach.
This is entirely consistent within the ``brick wall''
model, and yields non-zero results for the entropy
(which is the same for other techniques along these
lines).  
The divergent terms are properly accounted for,
and we discuss the finite terms in this case in
section~\ref{sec:finex}.
The other approach is ``extremalization before 
quantization''~\cite{ghosh97}.
This is more controversial.
It is already known~\cite{alwis95,kim96,cai96,ghosh95,mann98} 
that, before an inverse temperature is specified
(so that we are ``off-shell''), the divergences
in the ``brick wall'' model in this case 
are more severe than for a non-extremal black hole,
and cannot be renormalized away.
However, we shall show in section~\ref{sec:extremal}
that working ``on-shell'', with vanishing
temperature, all the divergent terms are
identically zero.
In addition, we argue in section~\ref{sec:massive}
that the finite terms in the entropy are
also vanishing in this approach.
This is in agreement with other methods of
calculating black hole entropy (for example,
a Hamiltonian approach~\cite{kiefer99}),
which give zero entropy for ``extremalization
before quantization'' and non-zero
entropy for ``extremalization after quantization''.
We would argue that the answers given by
the ``brick wall'' model are in fact
valid only ``on shell'', i.e. when at the end
the temperature has been fixed to be
the Hawking temperature, since
this is the only natural temperature 
at which to consider the quantum field.
That this outlook gives comparable answers
to other methods lends further weight to 
our approach.

Having shown that, by a suitable renormalization 
of the coupling constants in the effective gravitational
Lagrangian, a finite result for the entropy can be 
obtained, we then in section~\ref{sec:finite} proceed
to calculate and interpret this finite entropy.
This calculation cannot be performed analytically
for a general black hole space-time, so instead
we consider the simplest specific black hole geometry
in asymptotically adS space,
namely Schwarzschild-adS (S-adS).  
In this section we also restrict attention to
a massless scalar field in order to make the
integrals more tractable.  
The introduction of quantum field mass is not
expected to alter the qualitative nature of our
results.

One subtlety in the renormalization process is exactly
which logarithmic terms to discard. 
Since we can only take the logarithm of a dimensionless
quantity, it is necessary to renormalize away a 
term proportional to 
$\log (\epsilon /\Upsilon )$,
where $\Upsilon $ is some length scale, and 
$\epsilon $ is the co-ordinate distance of the 
``brick wall'' from the event horizon (not to be
confused with ${\tilde {\epsilon }}$, the proper distance
from the event horizon).  
The question is, what is the appropriate length scale
$\Upsilon $ in this case?  
We confirm that for a general choice of
length scale $\Upsilon $, in accordance with
the results of other authors~\cite{mann98}, the finite
entropy contains terms proportional to the logarithm of the
radius of the event horizon.  
Corrections to the Bekenstein-Hawking entropy of this
form have been found in other approaches to black hole entropy,
for example, using quantum geometry~\cite{kaul00}.
However, we also argue that there is a natural
choice of length scale for black holes in adS for which
these terms are absent.
For other black holes (not in adS), it remains an
open question as to the most appropriate choice of length scale.
Having isolated this logarithmic term ambiguity, our
focus is then the remaining finite entropy contribution.

For the Schwarzschild-adS black hole, a rather 
complex expression for the remaining quantum field entropy is obtained.
It is not readily apparent that this term arises
from the one-loop gravitational Lagrangian, so we firstly
consider the limit in which the radius of the event
horizon is much larger than the length scale
set by the negative cosmological constant.
In other words, we are considering a ``large''
black hole geometry in which the negative
cosmological constant has a great effect on the 
physics of the situation.
This is the limit of greatest interest, since in 
this limit the semi-classical approximation we are 
using is most valid, and, furthermore, the
black hole can be in stable equilibrium with
a thermal bath of radiation at the Hawking
temperature which is surrounding the event 
horizon~\cite{hawking83}.
In this limit we find that the remaining finite entropy is
entirely accounted for by terms coming from the 
one-loop effective Lagrangian.
Therefore, in this limit, we have an entirely
consistent picture.
The entropy of the black hole, including semi-classical
effects, arises from the one-loop effective theory.

We also consider the other limit, namely ``small'' black holes,
for which the radius of the event horizon is much smaller
than the length scale set by the cosmological constant.
In this case the entropy cannot be explained by the
effective Lagrangian, and dimensional considerations suggest
that it arises from a Lagrangian proportional to 
$R^{\frac {7}{2}}$,
where $R$ is the Ricci scalar of the geometry.
We would expect that for small black holes, higher-order or 
non-perturbative quantum gravity effects would be significant.
Therefore it is not surprising that our semi-classical
approach breaks down in this limit and we are led to 
corrections which involve fractional powers of the curvature. 
 
Next we consider the corresponding calculation
for a Reissner-Nordstr\"om-adS (RN-adS) black hole,
only in the limit in which the black hole is extremal.
We follow a ``extremalization after quantization'' approach,
namely the calculation is performed for a non-extremal
black hole and then the limit in which the black hole
becomes extremal is taken right at the end.  
The expression for the finite entropy is considerably
simpler in this case, and is easily explained
as arising from the one-loop Lagrangian, 
apart from the logarithmic terms, as discussed 
previously.

If we had taken the ``extremalization before quantization''
approach, for a massless field we show very easily that
the entropy (both finite and divergent terms) must vanish.
Therefore we close section~\ref{sec:finite} by showing
that the same is true if we consider a massive quantum
scalar field.
The integral required is intractable for a general 
extremal RN-adS black hole, so we focus on
one set of values for the black hole parameters,
in which the calculation is simplest.
A summary of our results and conclusions can be found in
section~\ref{sec:conc}.

\section{Renormalization of quantum scalar field entropy}
\label{sec:renorm}

We now consider the entropy of a quantum scalar
field on a general, static, spherically symmetric, black hole
background.
Firstly, we shall review the standard calculation of
the entropy using the WKB approximation, 
following~\cite{thooft85}.

\subsection{Entropy calculation using the WKB approximation}
\label{sec:WKB}

We use standard Schwarzschild-like co-ordinates, so that
the metric of our background geometry takes the form
\begin{equation}
ds^{2}= -\Delta (r) e^{2\delta (r)} dt^{2} 
+\Delta ^{-1} dr ^{2} + r^{2} \, d\theta ^{2}
+r^{2} \sin ^{2} \theta \, d\phi ^{2} .
\label{eq:metric}
\end{equation}
Here, and throughout this paper, the metric
has signature $(-,+,+,+)$ and we use
geometric units in which 
$G=\hbar =c =k_{B}=1$, where $k_{B}$ is
Boltzmann's constant,
except in section~\ref{sec:renormalize}, where
Newton's constant will be explicitly retained.
The metric function  $\Delta (r)$ is given by
\begin{equation}
\Delta (r) = 1 -
\frac {2m(r)}{r} - \frac {\Lambda r^{2}}{3}.
\end{equation}
We have included a cosmological constant $\Lambda $,
which we will usually assume is negative, so that
the black hole geometry approaches anti-de Sitter 
space far from the event horizon (this means
that the function $m(r)$ approaches a constant
as $r\rightarrow \infty $ at infinity).
Inclusion of a negative cosmological constant
means that the contribution to the entropy 
far from the black hole is finite.
This will enable us to concentrate on the 
divergences in the entropy arising from
the behaviour of the quantum field close
to the event horizon, which are our
main interest in this section.
Using a negative rather than a positive
cosmological constant ensures further that
there is no cosmological horizon to complicate
the issue~\cite{cai96}.  
We anticipate that our results concerning
the renormalization of the divergences
arising from close to the event horizon
would not be
significantly altered if the cosmological
constant were zero or positive.
At this stage we should comment that the 
metric~(\ref{eq:metric}) is in fact the most general
for the geometry outside the
event horizon (where $\Delta =0$), when
the black hole is static and spherically
symmetric, since we do not preclude the
possibility that $r=0$ at the event horizon.
Hereafter we shall denote the radius of the event
horizon as $r_{h}$, bearing in mind that
$r_{h}$ may be zero, as is the case for some
extremal black holes in string theory~\cite{alwis95,ghosh94}.

The field equation for a minimally coupled
quantum scalar field $\Phi $ of mass $\mu $
on the background~(\ref{eq:metric}) is
\begin{equation}
\nabla _{\nu }\nabla ^{\nu } \Phi -\mu ^{2}\Phi =0 .
\end{equation}
If $\Phi $ has the separable form
\begin{equation} 
\Phi (t,r,\theta ,\phi )=e^{-iEt} 
Y_{lm}(\theta , \phi) f(r),
\end{equation}
where $Y_{lm}(\theta, \phi )$ is the usual spherical
harmonic function,
then the equation for $f(r)$ reads
\begin{equation}
e^{-\delta } \left(
r^{2}e^{\delta } \Delta f' \right) '
+\left[
\frac {r^{2}E^{2}}{\Delta } e^{-2\delta } 
-\mu ^{2}r^{2} -l(l+1) \right] f =0 .
\label{eq:feqn}
\end{equation}
In order to use the WKB approximation, we define an $r$-dependent
radial wave number $K_{El}(r)$ by
\begin{equation}
K^{2}_{El}(r)= \frac {E^{2}}{\Delta ^{2}} e^{-2\delta }
-\frac {\mu ^{2}}{\Delta } -\frac {l(l+1)}{r^{2}\Delta }
\label{eq:Kdef}
\end{equation}
whenever the right-hand-side is positive, and $K_{El}=0$ otherwise.
By defining a new function ${\tilde {f}}(r)$ by
\begin{equation}
{\tilde {f}}(r)= {\sqrt {h(r)}} f(r),
\end{equation}
where $h(r)=e^{\delta } r^{2} \Delta $,
the equation~(\ref{eq:feqn}) now has the WKB form:
\begin{equation}
{\tilde {f}}''+
\left[ K^{2}_{El}(r) +\frac {1}{4h^{2}} h'^{2}
-\frac {1}{2h} h'' \right] {\tilde {f}} =0.
\end{equation}
The WKB approximation is valid when $K_{El}^{2}$ is the
dominant term in the square brackets, and will therefore
be a particularly good approximation for waves with large  
$E$ or $l$.
We will regularize our calculations by introducing
infra-red and ultra-violet cut-offs, so that
$\Phi =0$ when $r=r_{h}+\epsilon $ or $r=L$,
where $\epsilon \ll 1$ and $L\gg r_{h}$.
Therefore we are placing a ``brick wall'' a small distance
above the event horizon of the black holes.
We are also putting the whole system (black hole
and quantum field) in a large box, although
we shall see shortly that we can in fact take
the box to be infinitely large in the
situation where we have a negative cosmological constant. 
The number of radial waves $n_{El}$ satisfies the
semi-classical quantization condition
\begin{equation}
\pi n_{El} =\int _{r_{h}+\epsilon }^{L}
K_{El}(r) \, dr ,
\end{equation}
and then $N_{E}$, the total number of modes with
energy less than or equal to $E$ is given by
\begin{equation}
\pi N_{E} = \int (2l+1) \pi n_{El} \, dl ,
\end{equation}
where we have summed over the $(2l+1)$ modes
having different values of $m$ for the same $l$ and $E$,
and the ranges of all integrals are restricted by 
the fact that $K_{El}$ vanishes if the right-hand-side 
of~(\ref{eq:Kdef}) is negative. 
 
The free energy of the quantum scalar field at inverse
temperature $\beta $ is given by
\begin{equation}
e^{-\beta F} =\sum _{\rm {modes}} e^{-\beta E}
=\prod _{n_{El},l,m} \frac {1}{1-e^{-\beta E}},
\end{equation}
from which we deduce that
\begin{eqnarray}
F & = &
\frac {1}{\beta }\sum _{n_{El},l,m} \log \left( 1-e^{-\beta E} \right)
\nonumber \\
& = & 
\frac {1}{\beta }
\int dl \, (2l+1) \int dn_{El} \, \log \left( 1-e^{-\beta E} 
\right) 
\nonumber \\
& = & 
-\frac {1}{\beta }
\int dl \, (2l+1) \int d(\beta E) \frac {n_{El}}{e^{\beta E}-1}
\nonumber \\
& = & 
-\frac {1}{\pi } \int dl \,  (2l+1)
\int dE \frac {1}{e^{\beta E}-1}
\int _{r_{h}+\epsilon }^{L} dr \, K_{El}(r).
\end{eqnarray}
The $l$ integral can be performed explicitly to give
\begin{equation}
F= -\frac {2}{3\pi } \int \frac {dE}{e^{\beta E}-1}
\int _{r_{h}+\epsilon }^{L} dr
\frac {r^{2}}{\Delta ^{2}} e^{-3\delta }\left[
E^{2}-\mu ^{2} \Delta e^{2\delta } \right] ^{\frac {3}{2}} 
= -\frac {2}{3\pi } \int \frac {dE}{e^{\beta E}-1} I(E).
\label{eq:free}
\end{equation}
We close this subsection by considering the contribution 
to~(\ref{eq:free}) for $r\gg 1$.
The leading order contribution is:
\begin{equation}
F\sim -\frac {2}{3\pi } \int \frac {dE}{e^{\beta E}-1}
\int ^{L} dr \frac {9e^{-3\delta }}{\Lambda ^{2}r^{2}}
\left[ E^{2}+\frac {1}{3} \mu ^{2} e^{2\delta } \Lambda r^{2}
\right] ^{\frac {3}{2}} .
\end{equation}
If the field is massless ($\mu ^{2}=0$), it is clear 
that this expression tends to zero as $L\rightarrow \infty $.
On the other hand, if the field is massive this is not so obvious.
However, in this case, the requirement that the argument
of the square root be positive bounds $E$ from below (since
$\Lambda <0$):
\begin{equation}
E^{2}> -\frac {1}{3} e^{2\delta } \Lambda r^{2} \mu ^{2}.  
\end{equation}
Therefore, when $r\gg 1$, this lower bound on $E$ tends to 
infinity, which means that the integral over $E$ tends to zero.
In either case, the presence of a cosmological 
constant is crucial.
If the field is massive and $\Lambda =0$, then the lower bound
on $E$ becomes $E^{2}\ge \mu ^{2}$ for $r\gg 1$, 
so the $E$ integral does not vanish in this case.
The contribution for $r\gg 1$ in the asymptotically flat
situation is well understood as the free energy of a quantum
field in flat space. 
However, it clarifies the situation for our purposes 
to include the negative cosmological constant
so that we do not have to consider this part of the free
energy any further.

\subsection{Contribution from near the event horizon}
\label{sec:horizon}

We now consider the contribution to the free 
energy~(\ref{eq:free}) from close to the event horizon. 
For the time being, we focus on the case when the
black hole is non-extremal, leaving discussion of the
extremal case until section~\ref{sec:extremal}.
Early work in this area can be found in 't Hooft's 
original article~\cite{thooft85}, and also~\cite{mann92},
where a general non-extremal black hole near-horizon
geometry was transformed into Rindler space.

For a non-extremal black hole, we may expand $\Delta $
close to the event horizon as
\begin{equation}
\Delta = \Delta '_{h}(r-r_{h}) +
\frac {1}{2} \Delta ''_{h}(r-r_{h})^{2}
  +O(r-r_{h})^{3},
\label{eq:deltah}
\end{equation}
where the subscript $h$ denotes a quantity evaluated at the
horizon $r=r_{h}$. 
Note that $\Delta _{h}'$ cannot be zero since the black hole
is non-extremal.
The other quantities in~(\ref{eq:free}) can be expanded 
similarly, so that the contribution to the
$r$ integral from close to the event horizon is:
\begin{eqnarray}
I(E) & = & 
\int _{r_{h}+\epsilon } dr \,
\frac {E^{3}r_{h}^{2} e^{-3\delta _{h}}}{\Delta _{h}'^{2}(r-r_{h})^{2}}
\left[
1+(r-r_{h}) \left\{
\frac {2}{r_{h}} -3 \delta _{h}' 
-\frac {\Delta _{h}''}{\Delta _{h}'} -\frac {3}{2}
\frac {\Delta _{h}'}{E^{2}} \mu ^{2} e^{2\delta _{h}} 
\right\}
\right] +O(1)
\nonumber \\
& = &
\frac {E^{3}r_{h}^{2} e^{-3\delta _{h}}}{\Delta _{h}'^{2}}
\left[
\frac {1}{\epsilon }-\left\{
\frac {2}{r_{h}} -3 \delta _{h}' 
-\frac {\Delta _{h}''}{\Delta _{h}'} -\frac {3}{2}
\frac {\Delta _{h}'}{E^{2}} \mu ^{2} e^{2\delta _{h}} 
\right\} \log \epsilon 
\right] +O(1).
\end{eqnarray}
In order to perform the $E$ integral, we require the following
standard formulae~\cite{gradshteyn} 
\begin{eqnarray}
\int _{0}^{\infty } dE \frac {E^{3}}{e^{\beta E}-1} & = &
\frac {\pi ^{4}}{15 \beta ^{4}}
\nonumber \\
\int _{0}^{\infty } dE \frac {E}{e^{\beta E}-1} & = & 
\frac {\pi ^{2}}{6\beta ^{2}}.
\label{eq:standard}
\end{eqnarray}
Then the contribution to the free energy~(\ref{eq:free}) becomes:
\begin{equation}
F = -\frac {2}{3\pi } 
\frac {r_{h}^{2}e^{-3\delta _{h}}}{\Delta _{h}'^{2}}
\left[
\frac {1}{\epsilon } \frac {\pi ^{4}}{15\beta ^{4}}
-\left\{ \frac {2}{r_{h}}  -3 \delta _{h}' 
-\frac {\Delta _{h}''}{\Delta _{h}'} \right\}
\frac {\pi ^{4}}{15 \beta ^{4}} \log \epsilon
+\frac {\pi ^{2}}{4\beta ^{2}} \mu ^{2}
\Delta _{h}' e^{2\delta _{h}} \log \epsilon \right] +O(1).
\end{equation}
The entropy $S$ is calculated from the free energy $F$ using
the relation
\begin{equation}
S=\beta ^{2} \frac {\partial F}{\partial \beta } .
\label{eq:Sdef}
\end{equation}
Finally, we then substitute in the value of $\beta $,
namely the inverse Hawking temperature, so that
\begin{equation}
\beta = \frac {4\pi }{\Delta _{h}'} e^{-\delta _{h}} .
\end{equation}
This gives the contribution to the entropy from near the
event horizon as
\begin{equation}
S=\frac {1}{\epsilon }\frac {r_{h}^{2}}{360} \Delta _{h}'
-\left[\frac {r_{h}^{2}}{360} \left( 
\frac {2\Delta _{h}'}{r_{h}} - 3\delta _{h}' \Delta _{h}'
-\Delta _{h}'' \right) -\frac {r_{h}^{2}\mu ^{2}}{12} \right]
\log \epsilon +O(1). 
\end{equation}
The calculation is completed by noting that the variable $\epsilon $
is dependent on our choice of co-ordinates.
A better cut-off to use is the proper distance of the brick wall
from the event horizon, ${\tilde {\epsilon }}$, which is given by
\begin{equation}
{\tilde {\epsilon }}=\int _{r_{h}}^{r_{h}+\epsilon } dr 
\, \Delta ^{-\frac {1}{2}}
=2\epsilon ^{\frac {1}{2}} \left( \Delta _{h}' 
\right) ^{-\frac  {1}{2}}
+O(\epsilon ^{\frac {3}{2}} ),
\label{eq:tepdef}
\end{equation}
in terms of which the entropy is now
\begin{equation}
S=\frac {r_{h}^{2}}{90} {\tilde {\epsilon }}^{-2}
-\left[\frac {r_{h}^{2}}{180} \left( 
\frac {2\Delta _{h}'}{r_{h}} - 3\delta _{h}' \Delta _{h}'
-\Delta _{h}'' \right) -\frac {r_{h}^{2}\mu ^{2}}{6} \right]
\log {\tilde {\epsilon }} + 
{\mbox {terms finite as }}{\tilde {\epsilon }}\rightarrow 0 .
\label{eq:entdiv}
\end{equation}
The form~(\ref{eq:entdiv}) of the divergent contributions 
to the entropy is in agreement with previous calculations
in the 
literature~\cite{solodukhin95,demers95,kim96,ghosh94,cai96}.

\subsection{Renormalization of the entropy}
\label{sec:renormalize}

Various mechanisms have been suggested in the literature
for regularizing the divergent contribution to the 
entropy~(\ref{eq:entdiv}), such as using Pauli-Villars 
regularization~\cite{demers95,kim96}.
Here we shall follow a very simple approach, and show that
the terms in the entropy which diverge as the cut-off
$\epsilon $ approaches zero can easily be absorbed in 
a renormalization of the coupling constants
in the one-loop gravitational action. 

As is well known, to one-loop the effective gravitational
Lagrangian includes quadratic curvature 
interactions (see, for example,~\cite{birrell}):
\begin{equation}
{\cal {L}}= 
\frac{1}{16\pi G_{B}} (R -2 \Lambda _{B}) +
\frac{1}{4\pi } \left[
a_{B} R^{2} + b_{B} R_{\rho \sigma }R^{\rho \sigma }
+c_{B} R_{\rho \sigma \tau \lambda } R^{\rho \sigma \tau \lambda }
\right]
\label{eq:effl}
\end{equation}
where $a_{B}$, $b_{B}$ and $c_{B}$ are the (bare) coupling
constants for the quadratic interactions and we have
included a (bare) Newton's constant $G_{B}$.
There is also a (bare) cosmological constant $\Lambda _{B}$,
but this term does not contribute to the entropy.

The classical gravitational entropy 
arising from the Lagrangian~(\ref{eq:effl}) 
for a black hole
having a bifurcate Killing horizon is~\cite{wald} 
\begin{equation}
S=-2\pi \int _{\Sigma } E^{\rho \sigma \tau \lambda }
n_{\rho \sigma } n_{\tau \lambda }
\label{eq:classent}
\end{equation}
where the integral is performed over the bifurcation surface
$\Sigma $, with binormal $n_{\rho \sigma }$, and 
$E^{\rho \sigma \tau \lambda }$ is the functional 
derivative of ${\cal {L}}$ with respect to 
$R_{\rho \sigma \tau \lambda }$ holding the metric and connection
constant.
Taking this functional derivative of~(\ref{eq:effl}) gives 
the following entropy~\cite{demers95}:
\begin{equation}
S=\int _{\Sigma } d^{2}x \, {\sqrt {h}} 
\left[ \frac {1}{8G_{B}} g^{\rho \tau } g^{\sigma \lambda } 
n_{\rho \sigma }n_{\tau \lambda }
+ 2a_{B}R +b_{B} R_{\rho \sigma } g^{\rho \sigma }_{\perp }
-c_{B} R^{\rho \sigma \tau \lambda } n_{\rho \sigma }
n_{\tau \lambda } \right] ,
\end{equation}
where $h$ is the determinant of the metric on the bifurcation surface 
and $g^{\rho \sigma }_{\perp }$ is the metric in the
normal sub-space to the bifurcation surface.
It is straightforward to compute the required curvature
components for the metric~(\ref{eq:metric}), giving the answer
\begin{eqnarray}
S & = & \frac{1}{4G_{B}} A_{h} -8 \pi a_{B} \left[
r_{h}^{2}\Delta _{h}'' +3 r_{h}^{2} \Delta _{h}' \delta _{h}'
+4 r_{h} \Delta _{h}' -2 \right]
\nonumber \\
& & 
-4 \pi b_{B} \left[ r_{h}^{2} \Delta _{h}'' + 
3 r_{h}^{2} \Delta _{h}' \delta _{h}' +2 r_{h}\Delta _{h}' \right]
+8 \pi c_{B} ,
\label{eq:Sextra}
\end{eqnarray}
where $A_{h}=4\pi r_{h}^{2}$ is the area of the event horizon. 

We can now compare this classical entropy with the divergent
terms in the entropy of the quantum field on this background, 
given by~(\ref{eq:entdiv}).
In section~\ref{sec:finite} we shall consider the finite
terms in the quantum entropy.
Comparing~(\ref{eq:entdiv}) and~(\ref{eq:Sextra}) it can be
seen that the divergences can be absorbed in a renormalization 
of the coupling constants, as follows:
\begin{eqnarray}
G_{B}^{-1} & \rightarrow & G_{B}^{-1} + \frac {1}{90\pi} 
{\tilde {\epsilon }}^{-2} -\frac {1}{6\pi }\mu ^{2} 
\log {\tilde {\epsilon }}
\nonumber \\
a_{B} & \rightarrow & a_{B}+ \frac{1}{720\pi } 
\log {\tilde {\epsilon }}
\nonumber \\
b_{B} & \rightarrow & b_{B}- \frac{1}{240\pi }
\log {\tilde {\epsilon }}
\nonumber \\
c_{B} & \rightarrow & c_{B}- \frac{1}{360\pi } 
\log {\tilde {\epsilon }} .
\label{eq:renorm}
\end{eqnarray}
Our conclusions are in agreement with the 
work of other 
authors~\cite{demers95,fursaev96,shimomura00}
who considered the renormalization of the
leading and sub-leading divergences in the 
entropy, but using different approaches.

It should be noted at this stage that we have been
a little cavalier in our treatment of the
$\log {\tilde {\epsilon }}$ terms. 
Strictly speaking, one can only take the 
logarithm of a dimensionless quantity, so we
should instead consider
$\log ({\tilde {\epsilon }}/\Upsilon )$,
where $\Upsilon $ is some length scale.
Although this does not affect the divergence 
properties of the entropy, it will
introduce a term proportional to $\log \Upsilon $ 
into the finite (renormalized) entropy.
This leads to a potential ambiguity in the
finite entropy, which will depend on our
choice of length scale $\Upsilon $, and
will be discussed further in the next section.

We stress that we have now accounted for
all the divergences in the quantum field entropy,
since the next order in $\epsilon $ contribution
will be of order $\epsilon $, and so vanish
as $\epsilon \rightarrow 0$.
Therefore only the finite contribution to the
entropy remains.
This is the focus of the present work, and
is discussed in section~\ref{sec:finite}.

\subsection{Extremal black holes}
\label{sec:extremal}
The corresponding entropy calculation for extremal black holes
is rather more complex than for non-extremal black holes.
The contribution to the entropy from infinity is unchanged,
so we concentrate on the contribution from close to
the event horizon.
In common with the literature, there are two approaches one can take.
The first is ``extremalization after quantization''~\cite{ghosh97}.
Here one proceeds exactly as for a non-extremal black hole (the 
calculation in the previous section) and only sets the black hole
to be extremal right at the end.  
This corresponds to considering a whole set of non-extremal black
holes, which are approaching an extremal black hole in a limiting
process.
The issue in this case is at what stage in the calculation one
should set the black hole to be extremal.
If we follow through the calculation in the non-extremal case,
including the (non-zero) temperature and proper distance
cut-off ${\tilde {\epsilon }}$, then the answer is~(\ref{eq:entdiv}).
Now setting the black hole to be extremal simplifies the result
to give
\begin{equation}
S=\frac {r_{h}^{2}}{90} {\tilde {\epsilon }}^{-2}
+\left[
\frac {\Delta _{h}''r_{h}^{2}}{180} +
\frac {r_{h}^{2} \mu ^{2}}{6} \right] \log 
{\tilde {\epsilon }} +
{\mbox {terms finite as }} {\tilde {\epsilon }} \rightarrow 0. 
\end{equation}
In essence, the divergent contributions to the entropy
are exactly the same as in the non-extremal case, and 
may similarly be absorbed in a renormalization 
of the coupling constants (with the earlier 
proviso concerning the logarithmic terms).
The only situation which is different is when $r_{h}=0$,
which is the case for some extremal stringy black holes 
(such as those considered in~\cite{alwis95,ghosh94}).
In this case, as observed in that article, the first
divergent term disappears and we are left 
only with the term proportional to 
$\log {\tilde {\epsilon }}$ (note that 
$\Delta _{h}'' r_{h}^{2}$ will not in general 
vanish in the limit $r_{h}\rightarrow 0$).
This means that for these black holes the entropy is not in
fact zero, but will have a contribution from the 
one-loop corrections to the gravitational Lagrangian.

The second approach is ``extremalization before quantization''~\cite{ghosh97}.
Now, we assume from the outset that the black hole is extremal.
The expansion of the metric function $\Delta $~(\ref{eq:deltah})
now reads
\begin{equation}
\Delta = \frac {1}{2} \Delta _{h}'' (r-r_{h})^{2}
+\frac {1}{6} \Delta _{h}'''(r-r_{h}) ^{3}
+\frac {1}{24} \Delta _{h}^{(iv)} (r-r_{h})^{4}
+\frac {1}{120} \Delta _{h}^{(v)} (r-r_{h})^{5}
+O(r-r_{h})^{6}.
\end{equation}
Here, in order to compute all the divergent contributions to the
entropy, it is necessary to continue the expansion for rather
more terms than was the case for the non-extremal black hole.
The contribution from close to the event horizon
to the $r$ integral in the expression for the
free energy~(\ref{eq:free}) is now rather complex.
It has the form
\begin{equation}
I(E)= A_{3} E^{3} \epsilon ^{-3} + A_{2} E^{3} \epsilon ^{-2}
+(A_{1} E^{3} + B_{1} E) \epsilon ^{-1} +
(A_{0} E^{3}+ B_{0} E ) \log \epsilon +O(1) ,
\end{equation}
where the $A$'s and $B$'s are numbers depending on $r_{h}$, 
the mass $\mu $ of the quantum field and
the metric functions $\Delta $ and $\delta $ and their derivatives 
at the event horizon, but independent of $E$.
The exact form of the $A$'s and $B$'s is not important
for our considerations here.
Using the standard integrals~(\ref{eq:standard})
the contribution to the free energy for a quantum
field at inverse temperature $\beta $ is
\begin{equation}
F=-\frac {2\pi ^{3}}{45 \beta ^{4}} \left[ 
A_{3} \epsilon ^{-3} + A_{2} \epsilon ^{-2} +A_{1} \epsilon ^{-1}
+A_{0} \log \epsilon \right]
-\frac {\pi }{9\beta ^{2}} \left[
B_{1} \epsilon ^{-1} + B_{0} \log \epsilon \right] +O(1). 
\end{equation}
This gives the contribution to the entropy to be
\begin{equation}
S=\frac {8\pi ^{3}}{45 \beta ^{3}} \left[
A_{3} \epsilon ^{-3} + A_{2} \epsilon ^{-2} + A_{1} \epsilon ^{-1}
+A_{0} \log \epsilon \right]
+\frac {2\pi }{9\beta } \left[ B_{1} \epsilon ^{-1}
+B_{0} \log \epsilon \right] +O(1).
\label{eq:Shorex}
\end{equation}
This result is in agreement with similar calculations in the 
literature~\cite{kim96,cai96,ghosh95}, giving the leading divergence 
as proportional to $\epsilon ^{-3}$.
For an extremal black hole, the proper distance 
from the event horizon of
any point outside the event horizon is infinite,
so we have to work only with the 
co-ordinate dependent cut-off $\epsilon $.
In~\cite{cai96}, a cut-off proportional to $\log \epsilon $
is used, which results in an exponential divergence
in the entropy.
This cut-off represents the proper distance from the 
event horizon, but with an infinite additive factor, 
the same for all points outside the event horizon, removed.
Since we shall shortly show that all the divergent terms
in the entropy vanish ``on shell'', we shall 
keep $\epsilon $ in our calculations.
 
However, at this stage we are still dealing with an ``off-shell''
entropy, since we have not specified $\beta $.
For an extremal black hole, the Hawking temperature vanishes,
so that $\beta ^{-1}=0$.
Substituting this into~(\ref{eq:Shorex}), regardless of 
the value of our cut-off, we see immediately that all
the divergent terms disappear. 
The entropy of the quantum scalar field on the extremal
black hole background, calculated via this approach, does
not require any regularization.

\section{Calculation of finite entropy}
\label{sec:finite}

\subsection{Massless scalar field}
\label{sec:massless}

We now turn to the calculation and interpretation of the
finite contributions to the entropy of the quantum scalar field.
For general black hole geometries, the integral for the free 
energy~(\ref{eq:free}) will not be tractable analytically,
but will require a numerical computation.  
Therefore, in this section, we consider only simple black hole
space-times, so that the integrals can be performed exactly.
We consider the Schwarzschild-anti-de Sitter and extremal
Reissner-Nordstr\"om-anti-de Sitter black holes, and, for 
simplicity, a massless scalar field.
Considering black holes which are in asymptotically anti-de 
Sitter space means that, as observed earlier, the contributions
to the free energy and entropy far from the black hole remain
finite.  
We have chosen the simplest examples of an extremal and
non-extremal black hole geometry in order
to simplify the algebra, and hopefully reveal the
essence of the situation.

For both these geometries, the metric function $\delta $
vanishes identically, so the integral for the free energy takes
the simpler form
\begin{equation}
F=-\frac {2}{3\pi } \int \frac {E^{3} \, dE}{e^{\beta E}-1} 
\int _{r_{h}+\epsilon }^{L} dr \frac {r^{2}}{\Delta ^{2}}
=-\frac {2\pi ^{3}}{45 \beta ^{4}} I_{r},
\end{equation}
where we have performed the $E$ integral and 
\begin{equation}
I_{r}= \int _{r_{h}+\epsilon }^{L} dr \frac {r^{2}}{\Delta ^{2}},
\label{eq:irdef}
\end{equation}
depends only on the geometry of the black hole, and not
on $E$ or $\beta $ (this is the simplification afforded
by considering only a massless scalar field).
The entropy, calculated via~(\ref{eq:Sdef})
is then
\begin{equation}
S=\frac {8\pi ^{3}}{45 \beta ^{3}} I_{r}.
\label{eq:Sexact}
\end{equation}
We now give the result of the computation of the 
integral $I_{r}$ in each case.

\subsubsection{Non-extremal black hole}
\label{sec:finnon}
For the Schwarzschild-adS geometry, the metric function $\Delta $
is given by
\begin{equation}
\Delta = 1 - \frac {2M}{r} -\frac {\Lambda r^{2}}{3},
\end{equation}
where $M$ is a constant.
We shall simplify the integral $I_{r}$ by using dimensionless
variables as follows:
\begin{equation}
r=r_{h}x, \qquad
\epsilon = {\hat {\epsilon }}r_{h}, \qquad
L={\hat {L}} r_{h} .
\end{equation}
Then $I_{r}$ can be written as
\begin{equation}
I_{r}=\frac {9}{\Lambda ^{2}r_{h}} 
\int _{1+{\hat {\epsilon }}}^{{\hat {L}}}
dx \, x^{4}(x-1)^{-2} (x^{2}+x+\xi )^{-2}
\label{eq:sadsint}
\end{equation}
where $\xi = -6M/\Lambda r_{h}^{3}$ is a dimensionless
positive parameter dependent on the geometry.
The integral can be performed exactly,
and includes terms proportional to 
${\hat {\epsilon }}^{-1}$ and
$\log {\hat {\epsilon }}$, which, as seen in 
section~\ref{sec:renormalize}, can be absorbed in an
appropriate renormalization of the coupling constants.
However, as mentioned in section~\ref{sec:renormalize},
there is an inherent ambiguity in this process because
of the need to introduce an arbitrary length scale
$\Upsilon $ into the logarithm term, so the 
renormalization of the logarithmic terms needs 
a little more careful attention.

At this stage in the calculation, we are
working with dimensionless variables, the
length scale we are using being the radius
$r_{h}$ of the event horizon.
Our divergent logarithmic term is therefore
\begin{equation}
\log {\hat {\epsilon }} = \log (\epsilon /r_{h})
=\log (\epsilon /\Upsilon ) -\log (r_{h}/\Upsilon ),
\end{equation}
where we have introduced a general length scale $\Upsilon $.
What remains in our finite entropy is now, of course, dependent
on our choice of length scale $\Upsilon $, and therefore
what we discard with the divergent terms.
As observed in~\cite{mann98}, for a general
$\Upsilon $, if we discard by renormalization the 
$\log (\epsilon /\Upsilon )$ term, we are left
in the finite entropy with a contribution proportional to
$\log (r_{h}/\Upsilon )$, in other words, proportional
to the event horizon radius.  
Such corrections to the black hole entropy have been 
found in other contexts, for example, in quantum geometry 
approaches~\cite{kaul00}.
However, it should be noted that where $\log r_{h}$
terms have arisen in the semi-classical entropy~\cite{mann98},
this has been for black holes in asymptotically flat space.
In that situation, there is only one length scale, the
event horizon radius, which is dependent on the black hole
geometry, and therefore using that as the length scale 
$\Upsilon $ does not seem entirely natural (although it does remove
the $\log r_{h}$ contribution to the entropy).
In our situation, one might argue that it would be more 
appropriate to consider discarding a term of the 
form $\log ({\tilde {\epsilon }}/\Upsilon )$ rather 
than $\log (\epsilon /\Upsilon )$ since
${\tilde {\epsilon }}$ is the proper distance
of the ``brick wall'' from the event horizon, whereas
$\epsilon $ is a co-ordinate dependent quantity
(using ${\tilde {\epsilon }}$ rather than $\epsilon $ 
makes no difference in asymptotically flat space).
Ignoring numerical and non-dimensional geometric factors,
using~(\ref{eq:tepdef}), 
\begin{equation}
{\tilde {\epsilon }} ^{2}\propto 
\epsilon l^{2} r_{h}^{-1}={\hat {\epsilon }} l^{2}, 
\end{equation}
where $l={\sqrt {-3/\Lambda }}$ is the
length scale set by the cosmological constant.
Therefore, renormalizing away the term
$\log ({\tilde {\epsilon }}/\Upsilon )$ will
leave a contribution to the finite entropy proportional to
$\log (l/\Upsilon )$ rather than $\log (r_{h}/\Upsilon )$.
Here, in adS, there are two length scales, $r_{h}$ and $l$. 
It would seem reasonable to suggest that for 
black holes in adS, the most natural length scale 
is that set by the cosmological constant $\Lambda $, especially
as $\Lambda $ is a constant in the gravitational Lagrangian
of the theory, and, further, does not get renormalized
in the renormalization of the entropy (see 
section~\ref{sec:renormalize}).
It could therefore be argued that the term
$\log ({\tilde {\epsilon }}/l)$ is in fact the most
natural term to renormalize away, for black holes
in adS, as we are considering here. 
This results in the absence of terms proportional
to the logarithm of the event horizon radius
in the semi-classical entropy. 
This is the strategy we shall follow in the remainder of the
paper, so that we are investigating the additional 
part of the finite entropy, which does not include
logarithmic terms (possibly containing $\log r_{h}$ terms)
arising due to the ambiguity in choosing an appropriate
length scale.  
However, we should stress that for general black holes 
{\em {not}} in adS, it is not clear what is an appropriate
choice of length scale $\Upsilon $, and that,
in general, there will be contributions to the 
finite entropy proportional to the logarithm of 
the event horizon radius.  

Therefore, we ignore those contributions to the
integral~(\ref{eq:sadsint}) which are
proportional to ${\hat {\epsilon }}^{-1}$ or
$\log {\hat {\epsilon }}$ and let 
${\hat {L}}\rightarrow \infty $ and
${\hat {\epsilon }} \rightarrow 0$ to give our
final finite answer, which is:
\begin{equation}
I_{r} =  \frac {9}{\Lambda ^{2}r_{h}}
(\xi +2)^{-3} (4\xi -1)^{-\frac {3}{2}}
{\cal {P}},
\end{equation}
where
\begin{eqnarray}
{\cal {P}} & = &
(2\xi +1)(4\xi -1)^{\frac {3}{2}} \log (\xi +2)
+\left( 4\xi ^{4} + 32 \xi ^{3} + 12 \xi ^{2} + 8\xi -2 \right)
\tan ^{-1} 
\left( \frac {{\sqrt {4\xi -1}}}{3}  \right) 
\nonumber \\  
 & & 
+\left( 5\xi ^{3}+4\xi ^{2} +\xi -1 \right)
{\sqrt {4\xi -1}}   .
\label{eq:calp}
\end{eqnarray}
The final result for the entropy can now be calculated 
from~(\ref{eq:Sexact}) using the value of the inverse temperature 
in this geometry, which is:
\begin{equation}
\beta = -\frac {12\pi }{\Lambda r_{h}} (\xi +2)^{-1},
\end{equation}
giving
\begin{equation}
S=-\frac {\Lambda r_{h}^{2}}{1080} 
\left( 4\xi -1 \right) ^{-\frac {3}{2}} {\cal {P}}.
\end{equation}
Thus we have a rather complicated expression for the 
exact (apart from logarithmic terms),
finite entropy of a massless quantum scalar field on
the Schwarzschild-adS black hole, given simply in terms
of the parameters governing the geometry.
The first term in ${\cal {P}}$~(\ref{eq:calp}) involves
a logarithm multiplied by a geometric factor which,
up to a constant factor independent of the geometry,
is the same as the coefficient of the $\log {\hat {\epsilon }}$  
term in the divergent contribution to the 
entropy~(\ref{eq:Sextra}).
We may therefore conclude that this term
comes from the classical entropy due to
a higher-curvature correction to the effective
action.
However, in this case the coupling constant
will contain (or be renormalized by) a factor
$\log (\xi +2)$, which is dependent on the 
geometry, whereas in~(\ref{eq:renorm}) 
the renormalization factors depended on
the cut-off only.  
This is, in fact, not as serious as it may appear,
since the geometric dependence is contained within
a logarithmic term.
This could easily be absorbed into the
$\log {\hat {\epsilon }}$ divergent term which
renormalizes the coupling constants.
Further, in any renormalization scheme, there are ambiguities
present in absorption (or otherwise) of finite
terms, in addition to the ambiguities discussed
earlier in this section concerning the logarithmic terms.  
Here we are primarily interested in the 
interpretation of the entropy of the quantum
field.
Therefore we shall not consider the term
proportional to $\log (\xi +2)$ any further,
since its origin is now understood,
and it may be regarded as having been absorbed
in a (finite) renormalization of the coupling constants
in the higher-curvature Lagrangian~(\ref{eq:effl}).

The remaining terms in ${\cal {P}}$ are not
so easily dealt with.
The extra entropy we have is:
\begin{eqnarray}
S_{{\rm {extra}}} & = &
-\frac {\Lambda r_{h}^{2}}{1080} 
\left( 4\xi -1 \right) ^{-\frac {3}{2}} \left[
\left( 4\xi ^{4}+32 \xi ^{3} +12 \xi ^{2} +8\xi -2 \right)
\tan ^{-1} \left( \frac {{\sqrt {4\xi -1}}}{3} \right)
\right.
\nonumber \\
& & \left.
+\left( 5\xi ^{3} +4 \xi ^{2} +\xi -1 \right) 
{\sqrt {4\xi -1}} \right].
\label{eq:Sbonus}
\end{eqnarray}
At first sight, $S_{{\rm {extra}}}$ seems to be divergent
as $\xi \rightarrow 1/4$.
Performing the calculation separately when $\xi =1/4$ gives
a finite answer however.  
In fact the expression~(\ref{eq:Sbonus}) has a regular limit
as $\xi \rightarrow 1/4$, as can be seen by using the Taylor
expansion
\begin{equation}
\tan ^{-1} \left( \frac {{\sqrt {4\xi -1}}}{3} \right)
=\frac {{\sqrt {4\xi -1}}}{3} 
-\frac {(4\xi -1)^{\frac {3}{2}}}{27} +
O(4\xi -1)^{\frac {5}{2}}.
\end{equation}
With this expansion, the term in square brackets becomes
$O(4\xi -1)^{\frac {3}{2}}$, giving a finite answer for 
$S_{{\rm {extra}}}$.
 
The extra contribution to the entropy, $S_{{\rm {extra}}}$
contains terms which are not proportional to the 
quantities arising from the higher curvature corrections
to the Lagrangian~(\ref{eq:effl}).
All these terms arise from inserting the lower limit
$x=1+{\hat {\epsilon }}$ in the integral 
$I_{r}$, and then letting 
$\epsilon \rightarrow 0$.

In order to interpret these terms, we first consider the 
corresponding calculation when $\Lambda =0$, so we
are dealing with the Schwarzschild geometry.
In this case the free energy is not finite as 
$L \rightarrow \infty $.
However, if we ignore those terms which diverge in this limit
(as they are simply the free energy of the scalar field in flat space),
together with the terms divergent in $\epsilon $ as
$\epsilon \rightarrow 0$ and any logarithmic terms, we are left with
\begin{equation}
S=-\frac {13}{810},
\end{equation}
which may be absorbed into the coupling constants in the
effective Lagrangian~(\ref{eq:effl}) in any number of ways,
for example:
\begin{equation}
c_{B} \rightarrow c_{B}+ \frac {13}{6480\pi}.
\end{equation}
Therefore the presence of the negative cosmological constant
is having a considerable effect on the entropy, although
the extra terms arise from contributions near the event horizon.

Including now the cosmological constant, there are two limits
of interest.
Firstly, consider the case when $-\Lambda r_{h}^{2} \gg 1$.
This corresponds to large black holes, whose event horizon radius
is large compared with the length scale $l={\sqrt {-3/\Lambda }}$
set by the cosmological constant.
In this situation, since
\begin{equation}
\xi = -\frac {6M}{\Lambda r_{h}^{3}} = 1- \frac {3}{\Lambda r_{h}^{2}}
\sim 1 
\end{equation}
the entropy contribution~(\ref{eq:Sbonus}) becomes, in this limit,
\begin{equation}
S_{\rm {extra}} = -\frac {\Lambda r_{h}^{2}}{360} 
\left[ \frac {\pi }{\sqrt {3}} +1 \right] ,
\end{equation}
which arises from the higher derivative terms in the 
Lagrangian~(\ref{eq:effl}), and may be absorbed
by an appropriate (finite) renormalization of the coupling
constants, such as:
\begin{equation}
b_{B}\rightarrow b_{B} - \frac {1}{720\pi } \left[
\frac {\pi }{\sqrt {3}} + 1 \right] ,
\end{equation}
where, again, there is more than one such renormalization 
which will work.

The second limit is when $-\Lambda r_{h}^{2}\ll 1$,
which corresponds to small black holes, having an event horizon
radius much smaller than the length scale $l$ set by the 
cosmological constant.
In this limit,
\begin{equation}
\xi \sim -\frac {3}{\Lambda r_{h}^{2}} = \frac {l^{2}}{r_{h}^{2}} 
\gg 1,
\end{equation}
and the dominant contribution to the entropy~(\ref{eq:Sbonus})
is
\begin{equation}
S_{\rm {extra}} =\frac {\pi }{720} 
\left( \frac {l}{r_{h}} \right) ^{3} .
\label{eq:Snonabsorb}
\end{equation}
The entropy terms which arise from the higher derivative
Lagrangian~(\ref{eq:effl}) are proportional to $(r_{h}/l)^{2}$,
so we cannot absorb~(\ref{eq:Snonabsorb}) into a renormalization
of the coupling constants in~(\ref{eq:effl}) without making
the coupling constants dependent on the geometry in a manner
rather more serious than the absorption of the logarithmic
terms (although the renormalization of the coupling
constants would involve only dimensionless parameters).
The question then is whether~(\ref{eq:Snonabsorb}) could
arise from a gravitational Lagrangian of a different kind.

We will argue only on dimensional grounds.
Suppose we have a gravitational Lagrangian which contains $n$th
powers of the curvature, for example
\begin{equation}
{\cal {L}}_{n}= \alpha _{n} R^{n} .
\end{equation}
Then the coupling constant $\alpha _{n}$ will have dimensions
of $[{\rm {length}}]^{2n-4}$.
If we then calculate the classical entropy~(\ref{eq:classent})
arising from this Lagrangian, we get a result of the form
(the notation as in section~\ref{sec:renormalize})
\begin{eqnarray}
S & = & 
\int _{\Sigma } d^{2}x \, {\sqrt {h}} \,
n\alpha _{n} R^{n-1}
\nonumber \\
& = & 
4 \pi n\alpha _{n} r_{h}^{4-2n}(-1)^{n-1} \left[
r_{h}^{2} \Delta _{h}'' + 3r_{h}^{2} \Delta _{h}'
\delta _{h}' + 4 r_{h} \Delta _{h}' -2 
\right] ^{n-1} .
\end{eqnarray}
We now specialize to the Schwarzschild-adS geometry,
with 
\begin{equation}
\xi \sim \frac {l^{2}}{r_{h}^{2}} \gg 1 .
\end{equation}
We obtain
\begin{equation}
S \sim \alpha _{n} r_{h}^{4-2n} ,
\end{equation}
where all the factors of $l$ have canceled.
It seems reasonable in this model to have $\alpha _{n}$
proportional to $l^{2n-4}$, since $l$ is the length scale
set by the coupling constant $-\Lambda $ of the theory.
Therefore
\begin{equation}
S \sim \left( \frac {l}{r_{h}} \right) ^{2n-4} ,
\end{equation}
up to numerical factors which are independent of the 
black hole geometry.
Comparing with~(\ref{eq:Snonabsorb}), we get
agreement only if $n=7/2$, in other words our
extra contribution to the entropy can only arise
from a gravitational Lagrangian of the form
\begin{equation}
{\cal {L}}_{n}= \alpha _{\frac {7}{2}} R^{\frac {7}{2}} .
\end{equation}

For general $\xi $, the expression for the extra
entropy~(\ref{eq:Sextra}) is sufficiently complex
that we cannot make specific predictions about its source.
However, we have seen that it is completely described 
(modulo the ambiguities in the
terms containing logarithms, including, potentially, a term
proportional to 
the logarithm of the event horizon radius)
by contributions from the one-loop effective
gravitational Lagrangian~(\ref{eq:effl}) when
the black hole is large.
This is in accordance with expectations, for when
the black hole is large, the semi-classical
approximation we are using is valid, and 
higher-order and non-perturbative corrections
to the gravitational Lagrangian should be
negligible.
Therefore, in this case, the entropy of the
quantum scalar field is completely accounted for.
Our inclusion of a negative cosmological constant
is particularly pertinent in this situation,
as large black holes in anti-de Sitter space
have a stable Hartle-Hawking state, whereas
this state is unstable in asymptotically flat
space.

Our calculations indicate that this is not the case
for small black holes.
However, it should be said that the semi-classical
approximation will break down in this limit, 
so we should not be too surprised that 
the corrections to the gravitational Lagrangian
that are predicted contain unusual terms
with the curvature to a fractional power.
It will be necessary to appeal to quantum gravity
effects in order to get a consistent picture
in this case.

In this section, we have considered the 
simplest possible non-extremal black hole
in adS, namely Schwarzschild-adS.
However, we anticipate that our 
conclusions will apply to more general
black holes in models with a negative
cosmological constant.
For example, black holes have been 
found in ${\mathfrak {su}}(2)$
Einstein-Yang-Mills theory which possess
classically stable gauge field hair in 
adS~\cite{winstanley99}, precisely
when $l^{2}/r_{h}^{2} \ll 1$,
which is the limit in which our semi-classical
approach is most valid.
In this limit,
the geometries of these black holes have the form
of small perturbations of the Schwarzschild-adS
space-time. 
Therefore, to leading order in $l^{2}/r_{h}^{2}$,
the finite entropy of a massless quantum scalar
field on these black holes would be the same as
that we have calculated here for Schwarzschild-adS.
Of course, there would be sub-leading corrections due
to the additional structure of the black hole geometry,
and we hope to return to these matters in a subsequent
publication.

\subsubsection{Extremal black hole}
\label{sec:finex}

For the Reissner-Nordstr\"om-adS black hole, we shall
take the approach of ``extremalization after quantization''~\cite{ghosh97},
since putting $\beta ^{-1}=0$ in the general expression 
for the entropy~(\ref{eq:Sexact}) will give zero regardless
of the black hole geometry.  
For the RN-adS geometry, the metric function $\Delta $ 
has the form
\begin{equation}
\Delta = 1 - \frac {2M}{r} +\frac {Q^{2}}{r^{2}}
-\frac {\Lambda r^{2}}{3},
\end{equation}
where $Q$ is the charge. 
There are two horizons in this case, an outer (event) horizon 
at $r=r_{h}$ and an inner (Cauchy) horizon, at $r=r_{h}x_{-}$,
where $x_{-}\le 1$.
When $x_{-}=1$,  the inner and outer horizons merge and the
black hole is extremal.
As previously, we use dimensionless variables, and the integral
$I_{r}$ now takes the form
\begin{equation}
I_{r}=\frac {9}{\Lambda ^{2}r_{h}} 
\int _{1+{\hat {\epsilon }}}^{{\hat {L}}} dx \,
x^{6} \left( x-1 \right) ^{-2}
\left( x-x_{-} \right) ^{-2} \left(
x^{2}+(1+x_{-})x +\zeta \right) ^{-2},
\end{equation}
where the dimensionless parameter $\zeta $ is given by
\begin{equation}
\zeta = -\frac {3Q^{2}}{\Lambda r_{h}^{4} x_{-}} .
\end{equation}
Although this integral can be performed exactly, it is 
considerably more complicated than for Schwarzschild-adS.
However, matters are simplified because we are interested
solely in the extremal limit.
Using the dimensionless quantities, and substituting for
the inverse temperature $\beta $, which is now
\begin{equation}
\beta = -\frac {12\pi }{\Lambda r_{h}}
(1-x_{-})^{-1} (2+x_{-}+\zeta )^{-1},
\end{equation}
the expression for the 
entropy~(\ref{eq:Sexact}) takes the form:
\begin{equation}
S=-\frac {1}{9720} \Lambda ^{3} r_{h}^{3} (1-x_{-})^{3}
(2+x_{-}+\zeta )^{3}I_{r}.
\label{eq:SRN}
\end{equation}
Therefore we are concerned only with those terms in $I_{r}$
which remain when multiplied by $(1-x_{-})^{3}$ and the
limit $x_{-}\rightarrow 1$ taken.
As previously, we shall ignore those terms which vanish 
as $L\rightarrow \infty $ or $\epsilon \rightarrow 0$,
and also the known divergent terms in $\epsilon ^{-1}$
or $\log {\hat {\epsilon }}$ which can be absorbed in an appropriate
renormalization of the coupling constants.
As in the non-extremal case, there is some ambiguity
in the renormalization of the logarithmically divergent 
term $\log {\hat {\epsilon }}$, which is dependent
upon a choice of a length scale, and could give rise
to contributions to the finite entropy which are
proportional to the logarithm of the event horizon
radius~\cite{mann98}.
In this section we shall follow the strategy of 
section~\ref{sec:finnon}, and remove by renormalization
a term proportional to $\log {\hat {\epsilon }}$,
which will not leave any residual logarithmic terms in
the finite entropy.  
The remaining terms of interest are then:
\begin{eqnarray}
\frac {\Lambda ^{2}r_{h}}{9}
(1-x_{-})^{3} I_{r} &  = & 
\frac {x_{-}^{6}}{{\cal {A}}_{1}} 
+\frac {{\cal {A}}_{2}}{{\cal {A}}_{3}} \log (x_{-}-1) 
\nonumber \\ 
& & 
-\frac {{\cal {A}}_{4}}{{\cal {A}}_{5}} 
\frac {(1-x_{-})^{3}}{-4\zeta +1+2x_{-}+x_{-}^{2}}
\tanh ^{-1} \left( 
\frac {3+x_{-}}{{\sqrt {-4\zeta +1+2x_{-}+x_{-}^{2}}}}
\right) 
\nonumber \\ & &
+O(x_{-}-1) ,
\label{eq:rintRN}
\end{eqnarray}
where we have taken the limits $\epsilon \rightarrow 0$,
$L\rightarrow \infty $ and
retained the term involving $\tanh ^{-1}$ since it may be of
interest in the special case $\zeta = 1$.
In~(\ref{eq:rintRN}), the factors ${\cal {A}}_{i}$
are polynomials in $x_{-}$ and $\zeta $, which are non-zero
when either or both of $x_{-}$ or $\zeta $ are equal to unity.

The first term in~(\ref{eq:rintRN}) is finite as 
$x_{-}\rightarrow 1$, and, for this value of $x_{-}$, we have
\begin{equation}
{\cal {A}}_{1}=(\zeta +3)^{2}.
\end{equation}
The second term has a logarithmic singularity as 
$x_{-}\rightarrow 1$.
However, as has already been observed with other logarithmic terms,
the coefficient of $\log (x_{-}-1)$ in~(\ref{eq:rintRN}) 
is exactly the same as the coefficient of the $\log {\hat {\epsilon }}$
term in the integral.
Therefore we may absorb the $\log (x_{-}-1)$ term in 
an appropriate renormalization of the coupling constants.
The third term vanishes in the limit $x_{-}\rightarrow 1$
unless $\zeta =1$ also.
The case $x_{-}=\zeta =1$ can be dealt with in two ways:
firstly, we can set $\zeta =1$ for all $x_{-}$ and
then let $x_{-}\rightarrow 1$ subsequently; or secondly
we can set $\zeta = x_{-}$ and then let
$x_{-}\rightarrow 1$.
Proceeding either way, the third term in~(\ref{eq:rintRN})
vanishes in the limit $x_{-}\rightarrow 1$.

We may now take the limit $x_{-}\rightarrow 1$ in~(\ref{eq:rintRN})
and substitute in the expression for the entropy~(\ref{eq:SRN}).
The answer is:
\begin{equation}
S_{{\rm {extra}}} = -\frac {1}{1080} \Lambda r_{h}^{2}
(\zeta +3) 
=\frac {1}{360} \left[ \frac {Q^{2}}{r_{h}^{2}}-\Lambda 
r_{h}^{2} \right],
\label{eq:SbonusRN}
\end{equation}
where we have re-introduced the original geometric parameters.
The final expression for the entropy for an extremal RN-adS black 
hole is rather more simple than the corresponding 
expression~(\ref{eq:Sbonus}) for Schwarzschild-adS.
Furthermore, the expression~(\ref{eq:SbonusRN}) is exactly
proportional to the entropy coming from the
$R_{\rho \sigma }R^{\rho \sigma }$ term in the effective
Lagrangian~(\ref{eq:effl}).
Therefore the entropy in this case comes purely from
higher-order corrections to the gravitational 
Lagrangian, and may be absorbed in a suitable
redefinition of the coupling constants:
\begin{equation}
b_{B}\rightarrow b_{B} - \frac {1}{2880\pi } .
\label{eq:renormRN}
\end{equation} 
It is perhaps surprising that our expression for the 
entropy~(\ref{eq:SbonusRN}) is so much 
simpler than that for Schwarzschild-adS,
and that the simple renormalization~(\ref{eq:renormRN})
works for all values of $r_{h}$ in this case.
However, we are dealing in this situation
with an extremal black hole, which has zero
temperature, and therefore there are no
thermal excitations of the quantum field modes.
In the case of Schwarzschild-adS, we considered
the limit of large black holes, since it is
in this limit that the back-reaction 
on the geometry due to Hawking radiation is
expected to be small, and, further, the
Hartle-Hawking state representing the thermal
equilibrium between the black hole
and the thermal heat bath surrounding it is 
stable.
Since the temperature is zero for an
extremal black hole, this is not an issue,
and so our approach is valid for all values 
of $l/r_{h}$.

\subsection{Massive scalar field}
\label{sec:massive}

We close this section by considering a massive rather than
a massless scalar field, for an extremal black hole.
We anticipate that the inclusion of a scalar field mass
in the calculation of entropy for a non-extremal black hole
will not significantly alter the qualitative results of 
section~\ref{sec:finnon}, 
but it does greatly increase the computational
complexity.  
The integral corresponding to $I_{r}$~(\ref{eq:irdef}) for
a massive quantum scalar field on an S-adS black hole is not 
readily tractable in general.
Accordingly, it also seems reasonable to suppose that the inclusion
of scalar field mass will also not change the qualitative results
for the entropy on an extremal black hole geometry calculated via
an ``extremalization after quantization'' approach.
It may, however, make a difference to the ``extremalization before
quantization'' calculation, which we consider in this section.

For a quantum scalar field of mass $\mu $, the free energy is
given by~(\ref{eq:free}), which gives, for an extremal 
RN-adS black hole
\begin{eqnarray}
F & = &
-\frac {6}{\pi \Lambda ^{2} r_{h}} \int
\frac {dE}{e^{\beta E}-1} 
\int _{1+{\hat {\epsilon }}}^{{\hat {L}}} 
dx \,
x^{6} (x-1)^{-4} (x^{2}+2x+\zeta )^{-2}
\nonumber \\
& & \times
\left[ E^{2} +\frac {\mu ^{2} \Lambda  r_{h}^{2} }{3}
x^{-2} (x-1)^{2} (x^2+2x+\zeta ) \right] ^{\frac {3}{2}} ,
\label{eq:yukint}
\end{eqnarray}
where we are using dimensionless variables as in the previous
subsection. 
The $r$ integral is not readily tractable for general $\zeta $,
so we shall consider only the simplest case, when $\zeta =1$.
Setting
\begin{equation}
{\hat {E}}=E{\sqrt {\frac {-3}{\mu ^{2}\Lambda r_{h}^{2}}}},
\qquad
{\hat {\beta }}=\beta \frac {{\sqrt {-\Lambda }}}{{\sqrt {3}}}
\mu r_{h},
\end{equation}
the integral~(\ref{eq:yukint}) has the form
\begin{equation}
F=-\frac {2}{3\pi } \mu ^{4} r_{h}^{3} \int
\frac {d{\hat {E}}}{e^{{\hat {\beta }}{\hat {E}}}-1}
\int _{1+{\hat {\epsilon }}}^{{\hat {L}}} dx \,
x^{6} (x-1)^{-4} (x+1)^{-4} \left[
{\hat {E}}^{2} - \frac {1}{x^{2}} (x-1)^{2} (x+1)^{2}
\right] ^{\frac {3}{2}} .
\label{eq:yuk1} 
\end{equation}
The $r$ integral can be performed exactly. 
Inserting the limit ${\hat {L}}\rightarrow \infty $
gives no contribution, since the argument of the square
root in~(\ref{eq:yuk1}) must be positive, so that
when $x\gg 1$, it must be the case that ${\hat {E}}\gg 1$, which 
will give a vanishing ${\hat {E}}$ integral.
We already know from section~\ref{sec:extremal} that 
any terms which diverge as ${\hat {\epsilon }}\rightarrow 0$
will not contribute to the entropy.
Therefore,
inserting the lower limit $x=1+{\hat {\epsilon }}$ gives,
ignoring any divergent terms,
\begin{eqnarray}
F & = & \frac {1}{144\pi } \mu ^{4} r_{h}^{3}
\int \frac {d{\hat {E}}}{e^{{\hat {\beta }}{\hat {E}}}-1}
\left\{
43 {\hat {E}}^{3} -20 {\hat {E}}
+\left( 48 + 72 {\hat {E}}^{2} \right)
\tan ^{-1} \left( \frac {{\hat {E}}}{2} \right)
\right.
\nonumber \\
& & \left.
+\left( 15 {\hat {E}}^{3} + 108 {\hat {E}} \right)
\left[ \log 2 
-\frac {1}{2} \log \left( 1+ 
\frac {2}{{\hat {E}}^{2}} \right) \right] 
\right\} +O({\hat {\epsilon }}) .
\label{eq:yuk2}
\end{eqnarray}
With the standard integrals~(\ref{eq:standard})
it can be seen that the terms containing just ${\hat {E}}^{3}$
or ${\hat {E}}$ give, after performing the ${\hat {E}}$ integral,
terms proportional to $\beta ^{-4}$ or $\beta ^{-2}$ respectively.
When we then calculate the entropy, via~(\ref{eq:Sdef}), we end
up with terms proportional to $\beta ^{-3}$ or $\beta ^{-1}$.
Since we have an extremal black hole, these terms will all 
vanish when we substitute the ``on shell'' temperature 
$\beta ^{-1}=0$.
Then~(\ref{eq:yuk2}) reduces to
\begin{equation}
F = \frac {1}{144\pi } \mu ^{4} r_{h}^{3}
\int \frac {d{\hat {E}}}{e^{{\hat {\beta }}{\hat {E}}}-1}
\left\{
\left( 48 + 72 {\hat {E}}^{2} \right)
\tan ^{-1} \left( \frac {{\hat {E}}}{2} \right)
-\frac {1}{2}
\left( 15 {\hat {E}}^{3} + 108 {\hat {E}} \right)
\log \left( 1+ 
\frac {2}{{\hat {E}}^{2}} \right) 
\right\} .
\end{equation}
We shall now consider the integrals in the above equation in turn.

Firstly,
\begin{eqnarray}
\int \frac {d{\hat {E}}}{e^{{\hat {\beta }}{\hat {E}}}-1}
(48+72 {\hat {E}}^{2}) \tan ^{-1} \left(
\frac {{\hat {E}}}{2} \right) 
& & \nonumber \\ & & \hspace{-5cm} 
=\frac {2\pi }{{\hat {\beta }}} 
\left[ 48 \int
\frac {du}{e^{2\pi u}-1} \tan ^{-1}
\left( \frac {u}{q} \right) +
\frac {288\pi ^{2}}{{\hat {\beta }}^{2}} 
\int \frac {du}{e^{2\pi u}-1} u^{2}
\tan ^{-1} \left( \frac {u}{q} \right) \right] ,
\label{eq:tanint}
\end{eqnarray}
where
\begin{equation}
u=\frac {{\hat {\beta }}{\hat {E}}}{2\pi },
\qquad
q=\frac {{\hat {\beta }}}{\pi }
\end{equation}
These are now standard integrals, which can
be written for general $q$ in terms of gamma functions.
However, we are interested only in values of $q\gg 1$.
Therefore we may simplify by using the asymptotic
form of the gamma function for large $q$.
This gives
\begin{eqnarray}
\int \frac {du}{e^{2\pi u}-1} \tan ^{-1}
\left( \frac {u}{q} \right) 
& = &  \frac {1}{12q} + O \left( \frac {1}{q^{3}} \right) ;
\nonumber \\
\int \frac {du}{e^{2\pi u}-1} u^{2} \tan ^{-1}
\left( \frac {u}{q} \right) 
 & = & \frac {1}{60q} + O \left( \frac {1}{q^{2}} \right)  .
\end{eqnarray}
Therefore~(\ref{eq:tanint}) is, to leading order, proportional
to $\beta ^{-2}$, and this will give vanishing contribution
to the entropy.
The only remaining term which could contribute to the
free energy is now
\begin{eqnarray}
\int \frac {d{\hat {E}}}{e^{{\hat {\beta }}{\hat {E}}}-1}
\left( 15 {\hat {E}}^{3} + 108 {\hat {E}} \right)
\log \left( 1+\frac {2}{{\hat {E}}^{2}} \right) 
& & \nonumber \\ & & \hspace{-5cm}
=\int \frac {dv}{e^{v}-1} \left(
\frac {15}{{\hat {\beta }}^{4}} v^{3} +
\frac {108}{{\hat {\beta }}^{2}} v \right)
\left[ \log \left( 1+\frac {v^{2}}{2{\hat {\beta}}^{2}} \right)
+\log {2{\hat {\beta }}^{2}} -\log v^{2} \right] ,
\end{eqnarray}
where  $v={\hat {\beta }}{\hat {E}}$.
In this form, it is clear that this remaining integral has
leading behaviour ${\hat {\beta  }}^{-2} \log {\hat {\beta }}$
for ${\hat {\beta }}\gg 1$, so again there is no contribution 
to the entropy as $\beta ^{-1}\rightarrow  0$.

We therefore conclude that the exact entropy, for a massive
quantum scalar field, is identically zero in this approach.

\section{Conclusions}
\label{sec:conc}
In this article we have calculated the entropy of
a quantum scalar field in a thermally excited state
outside the event horizon of a black hole using
the ``brick wall'' model of 't Hooft~\cite{thooft85}.
We considered spherically symmetric black holes in 
asymptotically anti-de Sitter (adS) space, which means that
the theory is infra-red convergent.
Ultra-violet divergences remain due to the infinite
number of modes close to the event horizon,
and we regulate these by using a cut-off a
proper distance ${\tilde {\epsilon }}$
away from the event horizon (we focus on 
non-extremal black holes).
We showed that the divergences can all be
absorbed into a suitable renormalization
of the coupling constants in the one-loop 
effective gravitational Lagrangian,
yielding a finite entropy.
This is in agreement with similar calculations
in the literature performed via different methods.

However, the renormalization of the logarithmic 
divergences is dependent upon the choice of 
a length scale, and can lead to a contribution
to the entropy which is proportional to
the logarithm of the event horizon radius.
Terms of this form have been found in other
approaches to black hole entropy.
However, the picture here is not entirely clear, and
one natural choice of the length scale does not 
give any term of this form. 
We have argued that for black holes in adS, this
choice of length scale, namely that set by the 
cosmological constant, is in fact the most
reasonable to consider for this theory.
However, we consider this to be an open question requiring further 
investigation, particularly for general black holes
not in adS, where it is not clear what might be an appropriate
choice of length scale.

We next considered the finite entropy
(modulo these logarithmic ambiguities), 
for the particular case of the Schwarzschild-adS
black hole and a massless quantum scalar field.
For large black holes, the entropy so produced
is precisely accounted for from the one-loop
Lagrangian.
We conjecture that this will also be the case
for more general black holes,
but leave this question for future
investigations.
However, this is not the case for small
black holes and there are indications
from dimensional arguments that
non-perturbative quantum gravity
corrections may be important in this limit.

For extremal black holes, there are
two approaches that can be taken.
The first is ``extremalization after 
quantization'', which gives the
same qualitative results as in the 
non-extremal case. 
In other words, both the divergent and
finite contributions (apart from logarithmic terms) to the entropy 
are non-zero and arise
from the effective Lagrangian.
The second approach, ``extremalization
before quantization'', yields an
``on-shell'' entropy which has no
divergences.
The finite entropy vanishes for a massless
scalar field, and also for a particular
black hole with a massive scalar field.
We anticipate that this result will
also apply to more general extremal 
black holes.

Although our understanding of black hole
entropy has greatly increased over the
nearly thirty years the concept has been around,
many open questions remain and its
microscopic origin is not completely
known.
The semi-classical approach
advocated in this paper is 
straightforward both computationally
and conceptually, and readily
produces finite answers.
Our results in this paper 
reveal an encouraging 
consistency in this approach,
apart from an ambiguity in logarithmic
terms, although we have argued that there
is a natural choice of length scale
which does not produce any contribution
to the entropy proportional to the 
logarithm of the event horizon radius.
Namely, the entropy of the 
quantum field on the classical
background (apart from these logarithmic terms) is entirely 
accounted for by the 
additional terms in the gravitational
Lagrangian that must be included
in order to renormalize the
semi-classical theory.
In the large black hole
regime where this theory applies,
no new terms arise due to this scenario.
We anticipate that this result
can be extended to more general black hole
geometries, and quantum fields
of non-vanishing spin.
In the present paper, we have considered 
only black holes in adS space, motivated 
partly by the calculational simplicity
afforded by the regularization of the 
contribution to the entropy from infinity
in this case.
We consider that the main qualitative
conclusions of this article will also apply
to black holes not in adS space.
However, the details will need to be studied
separately, particularly with regard to 
the choice of cut-off.
We hope to return to these questions
in the near future.

\section*{Acknowledgments}
We would like to thank Parthasarathi Mitra, Sergey Solodukhin
and Robert Mann for helpful comments.

\end{document}